\documentclass[12pt,preprint]{aastex}

\slugcomment{Not to appear in Nonlearned J., 45.}

\shorttitle{Contribution of Intervening Clouds to GRB's X-ray Column Density}
\shortauthors{J. Wang}

\begin{document}


\title{Evidence of Contribution of Intervening Clouds to  
GRB's X-ray Column Density}


\author{J. Wang\altaffilmark{1}}
\altaffiltext{1}{National Astronomical Observatories, Chinese Academy of Sciences}
\email{wj@bao.ac.cn}



\begin{abstract}
The origin of excess of X-ray column density with respect to optical extinction in Gamma-ray bursts (GRBs) 
is still a puzzle. A proposed explanation of the excess is the photoelectric absorption due to the 
intervening clouds along a GRB's line-of-sight. We here test this scenario by using 
the intervening \ion{Mg}{2} absorption as a tracer of the neutral hydrogen column density of the intervening clouds.
We identify a connection between large X-ray column density (and large column density ratio of $\mathrm{\log(N_{H,X}/N_{HI})}\sim0.5$) 
and large neutral hydrogen column density probed by the \ion{Mg}{2} doublet ratio (DR).  
In addition, GRBs with large X-ray column density (and large ratio of $\mathrm{\log(N_{H,X}/N_{HI})}>0$) tend to have multiple 
saturated intervening absorbers with $\mathrm{DR<1.2}$.  
These results therefore indicate an additional contribution of the intervening system to the 
observed X-ray column density in some GRBs, although the contribution of the host galaxy alone cannot be 
excluded based on this study.

\end{abstract}





\section{Introduction}

Gamma-ray bursts (GRBs) are the most powerful explosions that occur throughout the universe
from local to extremely high redshift (e.g., Salvaterra et al. 2009; Tanvir et al. 2009).
The majority of GRBs show a long-duration ($T_{90}>2$s) and soft-spectrum in their prompt phase.
It is now generally accepted that these GRBs
originate from the death of young massive stars ($\geq25M_\odot$) according to the core-collapse model
(e.g., see reviews in Hjorth \& Bloom 2011 and Woosley \& Bloom 2006, and references therein).
The jet ignited in the core-collapse is believed to impact and shock the surrounding medium, which 
produces the GRB's afterglow through synchrotron radiation at a wide wavelength range from radio to X-ray 
(e.g., Meszaros \& Rees 1997; Sari et al. 1998).

Both X-ray and optical/near-infrared observations of afterglows provide an opportunity to study the metal abundance and 
column density in GRB's local environment (e.g., Savaglio et al. 2003; Butler et al. 2003; Prochaska et al. 2007).
Combining the spectral analysis in both bands suggests that the measured optical extinction is typically
smaller than the expectation inferred from the X-ray-derived column density of gas by 1-2 orders of magnitude,
although the column density is comparable to that in Galactic molecular clouds 
(e.g., Galama \& Wijers 2001; Stratta et al. 2004; Schady et al. 2007, 2010; Savaligo \& Fall 2004;
Greiner et al. 2011; Zafar et al. 2011; Kruhler et al. 2011; Campana et al. 2006; 2010; Watson et al. 2007). 
%In addition, the spectral analysis
%indicates that the Small Magellanic Cloud (SMC) extinction law without the 2175\AA\ abosorption feature 
%provides a better fit to the broad band afterglow spectral energy distribution as compared to the 
%extinction law of either Large Magellanic Cloud (LMC) or Milky Way (e.g., Stratta et al. 2004; Schady et al. 2010).

The large column density-to-dust ratio could be
explained by the destruction of small dust grains out to a radius of $\sim$10pc by the intense 
prompt emission of the GRB (e.g. Fruchter et al. 2001; Perna \& Lazzati 2001; Perna et al. 2003).
Campana et al. (2010) and Watson et al. (2007) claimed that the large 
ratio could be alternatively caused by either higher metallicity or stronger photoioniztion of hydrogen due to the
GRB's high energy emission. Schady et al. (2010), however, found a tight anti-correlation between
the column density-to-dust ratio and metallicity, which is in conflict with the expectation inferred from the
metallicity scenario. Kruhler et al. (2010) argued that the observed large ratio 
is mainly caused by two independent absorbers. One is a neutral absorber related to host galaxy that is located at a 
distance of $10^{2-3}$pc from the burst; the other is an ionized one that is in the vicinity of the burst.  
   
Thanks to the \it Swift\rm\ mission (Gehrels et al. 2004) and prompt follow-up observations in the optical band, 
a large GRB sample has recently shown that there is 
a correlation between the intrinsic X-ray column density and redshift (e.g., Campana et al. 2010; Behar et al. 2011;
Starling et al. 2013). In addition, Campana et al. (2012) and  Watson \& Jakobsson (2012) show that the bursts with
high column density can occur at low-to-moderate redshift when highly extinguished events are included. There 
is, however, a lack of bursts with low X-ray column density at high redshift. The correlation motivates various authors to argue that 
the X-ray column density excess could be explained by the photo-electric absorption due to diffuse intergalactic medium (IGM)
or intervening absorbing clouds, even though the intrinsic absorption from a host galaxy is required for nearby GRBs
(e.g., Behar et al. 2011; Starling et al. 2013; Campana et al. 2012). On the contrary, Watson \& Jakobsson (2012) argued that 
the intervening clouds alone seem to be insufficient to account for the large column density found in high-z GRBs.

In this paper, we present a piece of evidence supporting the conclusion that the observed X-ray column density excess is at least partially 
caused by the cumulative effect that results from the intervening systems in the line-of-sight of a GRB. Because of the poor spectral resolution
of current X-ray observations, the foreground absorption of \ion{Mg}{2}$\lambda\lambda2976, 2803$ doublet
is used as a tracer of the intervening systems. The paper is organized as follows. 
The sample selection is described in \S2. \S 3 then compares 
the intervening \ion{Mg}{2} absorption with the intrinsic X-ray column density to reveal the 
contribution from the intervening absorption systems.

\section{\ion{Mg}{2} Absorption Sample}

%% In a manner similar to \objectname authors can provide links to dataset
%% hosted at participating data centers via the \dataset{} command.  The
%% second curly bracket argument is printed in the text while the first
%% parentheses argument serves as the valid data set identifier.  Large
%% lists of data set are best provided in a table (see Table 3 for an example).
%% Valid data set identifiers should be obtained from the data center that
%% is currently hosting the data.
%%
%% Note that AASTeX interprets everything between the curly braces in the 
%% macro as regular text, so any special characters, e.g. "#" or "_," must be 
%% preceded by a backslash. Otherwise, you will get a LaTeX error when you 
%% compile your manuscript.  Special characters do not 
%% need to be escaped in the optional, square-bracket argument.

Thanks to the rapid-response spectroscopy of high-z GRB's optical afterglows,
the absorption of \ion{Mg}{2}$\lambda\lambda$2796, 2803 doublet is a commonly used transition in studying the intervening absorbing clouds in 
the line-of-sight of a GRB, both
because of the relative high abundance of Mg and because of the high strength of the 
transition (e.g., Prochter et al. 2006; Prochaska et al 2007; Tejos et al. 2009; Vergani et al. 2009; Cucchiara et al. 2012).

We compile a sample of all available \it Swift\rm\ GRBs with intervening \ion{Mg}{2} doublet absorption reported in literature.
No cut on the intervening \ion{Mg}{2}$\lambda2796$ equivalent width (EW) is used in our sample compilation.
We further require that the measured rest-frame EWs of both lines in the \ion{Mg}{2} doublet are available for each intervening absorber.
The sample finally contains 29 GRBs, and is tabulated in Table 1, along with the references. 
All the errors reported in the table correspond to the 1$\sigma$ significance level after taking into account the 
proper error propagation. 

For each GRB, Columns (1) and (2) list
the GRB identification and measured redshift of the GRB, respectively. The redshifts of the identified intervening \ion{Mg}{2} absorption systems
are tabulated in Column (3). Columns (4) and (5) present the EW of \ion{Mg}{2}$\lambda2796$ and the corresponding 
\ion{Mg}{2} doublet ratio $\mathrm{DR=EW(2796)/EW(2803)}$, respectively. 
For each afterglow with multiple intervening \ion{Mg}{2} absorption systems, 
Columns (4) and (5) list the values of the intervening absorber with the lowest DR value.  
The corresponding redshift of the absorber is marked in boldface type in Column (3).   
In these afterglows, we further check if there are multiple intervening absorbers with $\mathrm{DR<1.2}$ in each line-of-sight, 
and mark the corresponding redshifts with an asterisk in Column (3).  
The intrinsic X-ray 
column density $\mathrm{N_{H,X}}$ is listed in Column (6) for each of the GRBs. The column density is obtained from the X-ray energy spectrum in the 
\it Swift\rm\ XRT Photon Counting (PC) mode\footnote{The data and spectral analysis results can be found in the UK Swift Science Data 
Center repository at http://www.swift.ac.uk/xrt\_spectra (Evans et al. 2009).}. The spectrum taken in 
the Windowed Timing mode, if any, is not adopted to avoid the possible spectral evolution at early time (e.g., Campana et al. 2007; Gendre et al. 2007).

In the current sample, the X-ray column density $\mathrm{N_{H,X}}$ ranges from $10^{21}$ to $10^{22.5}\ \mathrm{cm^{-2}}$. 
The sample reproduces the previously reported $\mathrm{N_{H,X}}$ versus $z$ correlation (see the citations in INTRODUCTION). 
The relationship is presented in Figure 1 for the current sample.  
By excluding the GRBs with upper limits of $\mathrm{N_{H,X}}$, a statistical test returns a Kendall's $\tau=0.409$, $Z$-value of 
2.81 and $P=0.005$, where $P$ is the probability that there is no correlation between the two variables. A few of 
GRBs with high column density and heavy optical extinction have been identified in the low-to-moderate redshift range (Campana et al. 2012;
Watson \& Jakobsson 2012). The heavy extinction most likely results in 
a selection bias against these bursts in the current sample. Watson \& Jakobsson (2012) stated that the 
redshifts of these bursts are mainly obtained from the observations of their host galaxies.  
%Among the bursts with high column density and high optical extinction, GRB\,080607 is the only one event that is 
%listed in our \ion{Mg}{2} absorption sample. The GRB has extinction of $A_{\mathrm{V}}\sim2.3$mag (see below). 
%A careful inspection on Figure 2 in Watson \& Jakobsson (2012) shows that this GRB slightly deviates from the $\mathrm{N_{H,X}}$ 
%versus $z$ correlation.   

Figure 2 shows the relation between $\mathrm{N_{H,X}}$ and optical extinction $A_{\mathrm V}$ for the current sample, when
the values of $A_{\mathrm V}$ are reported in literature. We list the values of $A_{\mathrm V}$  in column (8) in Table 1.
The dashed line shown in the figure marks the typical dust-to-gas ratio for the Local Group in the case with solar metallicity (Welty et al. 2012). 
One can see from the figure an evident excess of $\mathrm{N_{H,X}}$ with respect to the optical extinction, and 
a marginal increase of $\mathrm{N_{H,X}}$ with $A_{\mathrm V}$. The mean value of $\mathrm{N_{H,X}}/A_{\mathrm V}$ ratio is 
$\approx1.7\times10^{22}\ \mathrm{cm^{-2}\ mag^{-1}}$, which is highly consistent with the one reported in Covino et al. (2013).

%% In this section, we use  the \subsection command to set off
%% a subsection.  \footnote is used to insert a footnote to the text.


%% This section also includes several of the displayed math environments
%% mentioned in the Author Guide.

\section{Results and Discussions}

\subsection{MgII doublet absorption}
We use the intervening \ion{Mg}{2} doublet absorption as a tracer of the foreground absorption system to study the origin of the 
X-ray column density excess detected in GRBs.
The \ion{Mg}{2} absorption traces a wide range of neutral hydrogen column density $\mathrm{N_{HI}}$ from
$10^{16}$ to $10^{22}\mathrm{cm^{-2}}$ (corresponding to an EW(2796) from 0.02 to 10\AA, 
e.g., Churchill et al. 2000; Rao \& Turnshek 2000; Rigby et al. 2002). 
By using the expanded SDSS/HST sample of low-z Lyman-$\alpha$ absorbers, Menard \& Chelouche (2009) reveals a relationship
between mean $\mathrm{N_{HI}}$ and EW(2796), although the relationship has very large scatter.

It has been frequently pointed out that the EW of \ion{Mg}{2}$\lambda$2976 line is a better indicator 
of number of individual components and/or the velocity spread of the gas than an indicator of $\mathrm{N_{HI}}$ (e.g., Petitjean \& Bergeron 1990; 
Prochter et al. 2006; Gupta et al. 2009). It is noted that a low \ion{Mg}{2} DR close to 1 more directly 
reflects high column density than does EW(2796) (e.g., Gupta et al. 2009 and references therein). 
The DR of the \ion{Mg}{2} doublet theoretically ranges from 1.0 for completely saturated lines to
2.0 for unsaturated lines according to atomic physics. In the current sample, 
$\sim$80\% GRBs have \ion{Mg}{2} DRs within the range $\mathrm{1<DR<2}$ (within the errors).
After taking into account the errors, the remaining GRBs listed in
our sample have \ion{Mg}{2} DRs very close to the upper and lower theoretical limits.
The mean and median values of DR are 1.18 and 1.15, respectively,
which means there is a strong saturation for these intervening \ion{Mg}{2} absorption clouds.

If the X-ray column density excess was only attributed to the gas related to the host galaxy, the strength of intervening \ion{Mg}{2} absorption is expected 
to be physically unrelated with the X-ray column density.  Figure 3 presents the \ion{Mg}{2} DR versus $\mathrm{N_{H,X}}$ diagram.
A moderately strong anti-correlation can be identified from
the diagram: the larger the \ion{Mg}{2} DR, the smaller the X-ray column density will be.
A statistical test yields a Kendall's $\tau=-0.2586$ and a $Z$-value of 2.091 at a significance
level with a probability of null correlation of $P=0.0365$.  These coefficients are calculated
through the survival analysis to take account of the entries in which only upper limits of $\mathrm{N_{H,X}}$  can be obtained from observations.  
The anti-correlation suggests a positive relationship between the observed X-ray column density
and the column density traced by the intervening \ion{Mg}{2} absorptions. 
The afterglows with multiple intervening absorbers are shown by the blue points, and the afterglows with a single absorber by the 
red ones. The size of the blue points is scaled to the 
number of saturated \ion{Mg}{2} absorbers with $\mathrm{DR<1.2}$. 
It seems that, as expected, the line-of-sight with more saturated absorbers tends to correspond to higher X-ray column density.  

A Monte-Carlo simulation with 1,000 random experiments is performed to investigate the effect caused by the larger errors 
in both DR and $\mathrm{N_{H,X}}$. A random sample is established by assuming a Gaussian distribution for each measurement (except for the upper limits).
The Kendall's $\tau$ test is then carried out for each of the 1,000 random samples.   
Again, the tests are based on the survival analysis. Figure 4 shows the distribution of the simulated Kendall's $\tau$.
The distribution can be well described by a Gaussian function with a peak at $\tau=-0.23$ and a standard deviation of 0.06.    
A weighted sum of the calculated probability of each random sample yields a total probability of null correlation of 0.368, which indicates that 
the significance of the correlation degrades to the 1$\sigma$ significance level when the large errors are taken into account.

\subsection{$\mathrm{N_{HI}}$ from Ly$\alpha$ observations}
Covino et al. (2013) recently compared the X-ray column density with the 
neutral hydrogen column density measured from the
local Lyman-$\alpha$ absorption. The comparison shows that there is a relation between the 
two independent measurements for some events listed in their sample. Some outliers, however, can be 
clearly identified from their comparison. Furthermore, the authors claimed that the relation is improved
when the contribution of the intervening system is removed from the observed X-ray column density.
The contribution is empirically estimated from the intervening absorption model 
proposed in Campana et al. (2012). 

In order to additionally test the effect caused by the intervening system, the \ion{Mg}{2} DR is plotted in Figure 5 as a 
function of the ratio between the column density measured from the X-ray spectrum and that measured from the local
Lyman-$\alpha$ absorption. The host galaxy hydrogen column density ($\mathrm{N_{HI}}$) obtained from the Lyman-$\alpha$ observation 
is listed in Column (7) in Table 1 for each GRB. The data are mainly taken from Fynbo et al. (2009), 
except for GRB\,100219 (Thone et al. 2013). 
GRB\,060607 and GRB\,050908 are not shown in the plot because of their extremely
large ratios.  The estimated ratios are $\log(\mathrm{N_{H,X}/N_{HI})}=4.85\pm0.19$ and $<3.9$ for GRB\,060607 and GRB\,050908,
respectively. The corresponding \ion{Mg}{2} DRs are as small as $1.02\pm0.08$ and $1.32\pm0.03$.  
One can see from the figure that the GRBs with large $\mathrm{N_{H,X}/N_{HI}}$ ratio (i.e., $\log(\mathrm{N_{H,X}/N_{HI})}>0.5$) tend to have heavy line saturation with 
\ion{Mg}{2} DR close to 1. Since a small \ion{Mg}{2} DR caused by the saturation of \ion{Mg}{2}$\lambda2976$ line reflects a 
high neutral hydrogen column density, the trend suggests that the large X-ray column density observed in some GRBs could be 
attributed to an additional absorption caused by their intervening system.              
Similar to Figure 3, the line-of-sights with multiple (single) saturated absorbers are presented by the blue (red) points. The size of the blue point 
is again scaled to the number of saturated absorbers. Although there is only one saturated absorbers in the 
line-of-sight of GRB\,060607, the GRBs with $\log(\mathrm{N_{H,X}/N_{HI})}>0$ tend to be associated with more than one 
saturated absorber.

\subsection{Conclusions and Discussions}

The discrepancy between $\mathrm{N_{HI}}$ obtained from optical data and that from X-ray spectral analysis has been frequently
reported in previous studies (e.g., Starling et al. 2007; Schady et al. 2010). 
By using the \ion{Mg}{2} doublet absorption that is taken from afterglow optical spectroscopy as a 
tracer of neutral hydrogen column density,
the statistical study in this paper reveals a connection in which a GRB with high X-ray column density tends to be associated with 
large intervening neutral hydrogen column density. This connection
therefore indicates an additional contribution of the intervening system to the observed $\mathrm{N_{H,X}}$ in some GRBs on observational 
grounds, which, however, does not mean a firm exclusion that the high $\mathrm{N_{H,X}}$ values can be due to the host galaxy component alone.
The contribution to X-ray column density needs to be quantified for each component to properly explain observations, which is a hard task
at present because the current
X-ray spectral resolution is too low to identify individual absorption features.

The revealed connection between intervening hydrogen column density and $\mathrm{N_{H,X}}$ can naturally explain the observed correlation 
between $\mathrm{N_{H,X}}$ and redshift by including an extra X-ray absorption contributed by the intervening system. 
The scenario in which $\mathrm{N_{H,X}}$ is dominantly contributed by the intervening system has been proposed by many previous studies.
By doubling the number of intervening clouds that are derived from QSO studies, 
the simulation done by Campana et al. (2012) suggests that the intervening absorber scenario is a plausible 
explanation for the X-ray column density excess. With a more realistic model, Starling et al. (2013) proposed that
the absorption due to the cumulative effect of intervening Lyman-$\alpha$ clouds can equally explain the X-ray column density excess. 
An alternative possible explanation of the discrepancy is the photoelectric absorption due to 
diffuse cold or warm intergalactic medium (Behar et al. 2011; Starling et al. 2013). 
A further support for the contribution of the intervening systems may come from the fact that the relation between 
$\mathrm{N_{H,X}}$ and $\mathrm{N_{HI}}$ become tighter after subtracting to the X-ray values the mean contribution from intervening 
systems (see Covino et al. 2013). Under some conditions, the intervening absorbers can also affect the 
$\mathrm{A_v}$ determination and possibly contribute to the observed $\mathrm{N_{H,X}}$ versus $\mathrm{A_v}$ correlation found by
Watson et al. (2013) and Covino et al. (2013).

%The aforementioned scenario can also explain the observed $\mathrm{N_{H,X}}-A_{\mathrm{V}}$ correlation 
%(e.g., Watson et al. 2013; Covino et al. 2013). In fact,  the dust extinction is derived from the fitting of afterglow's spectral 
%energy distribution from X-ray to near-infrared, which is, in principle, affected by the dust in the foreground galaxy along the 
%line-of-sight of a GRB. Prochaska et al. (2007) argued that the dampled Lyman-$\alpha$ systems identified in GRB's afterglow spectra 
%probe the dense gas located in the inner few kpc of the foreground galaxy. A direct support of the additional X-ray absorption 
%contributed by the intervening system comes from the tight  correlation between the $\mathrm{N_{H,X}}$ related to host 
%galaxy and Ly$\alpha$-derived $\mathrm{N_{HI}}$.
%The correlation was identified by Covino et al. (2003) after subtracting the contribution of the intervening clouds from the observed $\mathrm{N_{H,X}}$.     

Watson \& Jakobsson (2012) argued that the intervening clouds seem to be insufficient to explain the 
column density excess after taking into account of the metallicity evolution and unjustified overabundance of 
intervening clouds in the GRB's line-of-sight. In fact, a much weaker overabundance was recently revealed in 
Cucchiara et al. (2012) by re-analyzing a large sample of intervening \ion{Mg}{2} absorption. 
Watson et al. (2013) recently proposed that the absorption by Helium in the host \ion{H}{2} region is responsible for 
most of the observed X-ray column density, after excluding other scenarios. Their argument is based on the 
$\mathrm{N_{H,X}}-A_{\mathrm V}$ ($\mathrm{N_{HI}}$) correlation and a  
change of $\mathrm{N_{H,X}}/A_{\mathrm V}$ ratio with redshift that is similar to cosmic metallicity evolution. 
A recent study shows that the evolution of $\mathrm{N_{H,X}}/A_{\mathrm V}$ is likely caused by 
the lack of bursts with small $\mathrm{N_{H,X}}$ at high redshift rather than the evolution of $A_{\mathrm V}$,
although the physical origin of the lack has not been determined at the current stage (Covino et al. 2013).  
This result seems to be inconsistent with the implication of the scenario proposed in Watson et al. (2013).

\acknowledgments

The author thanks an anonymous referee for his/her careful review and 
helpful suggestions for improving the manuscript. The author
thanks James Wicker for help with language.
The study is supported by the National Basic Research Program of China (grant
2009CB824800) and by the National Natural Science
Foundation of China under grant 11003022. We gratefully acknowledge funding
for \it Swift\rm\ at the University of Leicester from the UK Space Agency. This 
work uses data provided by the UK \it Swift\rm\ Science Data Center at 
the University.

\clearpage

%% the file, uncomment the \plotone, \plottwo, and
%% If you need a layout that cannot be achieved with \plotone or
%% \plottwo, you can invoke the graphicx package directly with the
%% \includegraphics command or use \plotfiddle. For more information,
%% please see the tutorial on "Using Electronic Art with AASTeX" in the
%% documentation section at the AASTeX Web site,
%% http://www.journals.uchicago.edu/AAS/AASTeX.
%%
%% The examples below also include sample markup for submission of
%% supplemental electronic materials. As always, be sure to check
%% the instructions to authors for the journal you are submitting to
%% for specific submissions guidelines as they vary from

%% This example uses \plotone to include an EPS file scaled to
%% 80% of its natural size with \epsscale. Its caption
%% has been written to indicate that additional figure parts will be
%% available in the electronic journal.

\begin{figure}
\epsscale{.80}
\plotone{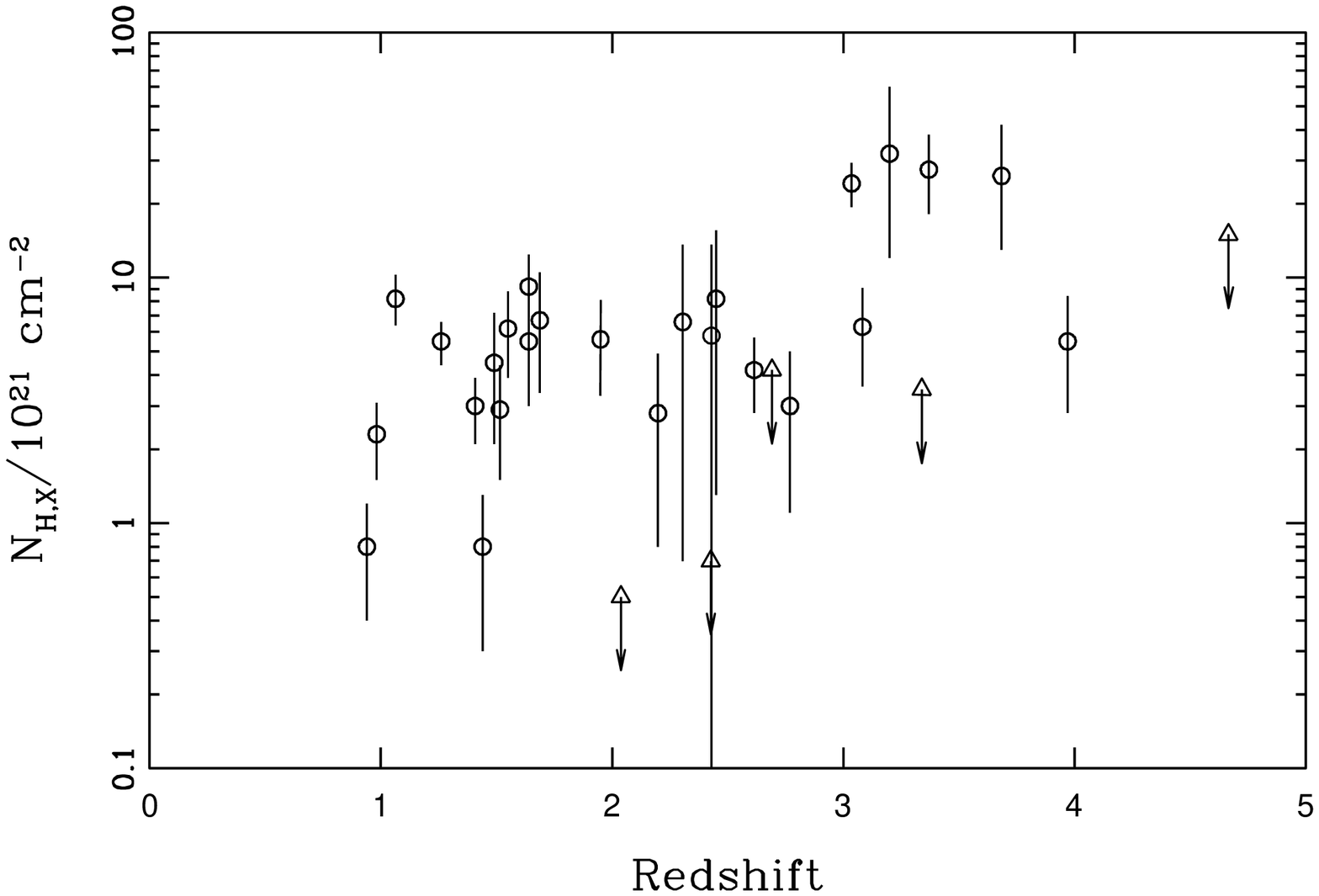}
\caption{Redshift evolution of the X-ray column density ($\mathrm{N_{H,X}}$) for the 29 GRBs listed in the 
current sample.  The GRBs with only the upper limit of  $\mathrm{N_{H,X}}$ are marked by the open 
triangles and arrows. The errorbars correspond to the 1$\sigma$ significance level.\label{fig1}}
\end{figure}

\begin{figure}
\epsscale{.80}
\plotone{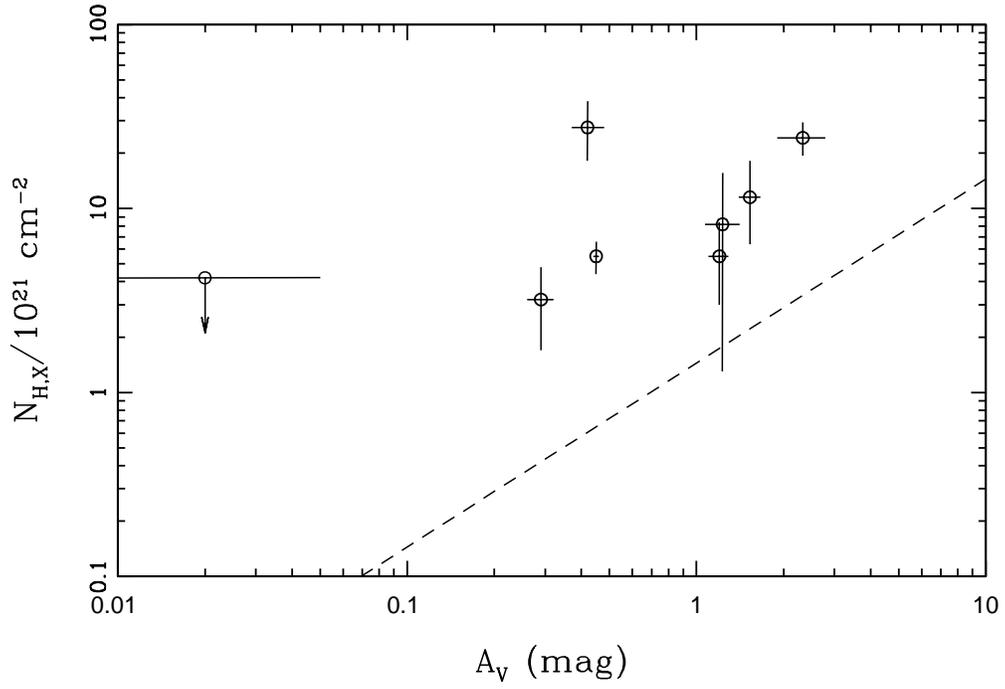}
\caption{X-ray column density $\mathrm{N_{H,X}}$ plotted against dust extinction $A_{\mathrm V}$. The dashed line shows the typical 
value of dust-to-gas ratio for the environment of the Local Group. \label{fig2}}
\end{figure}

\begin{figure}
\epsscale{.80}
\plotone{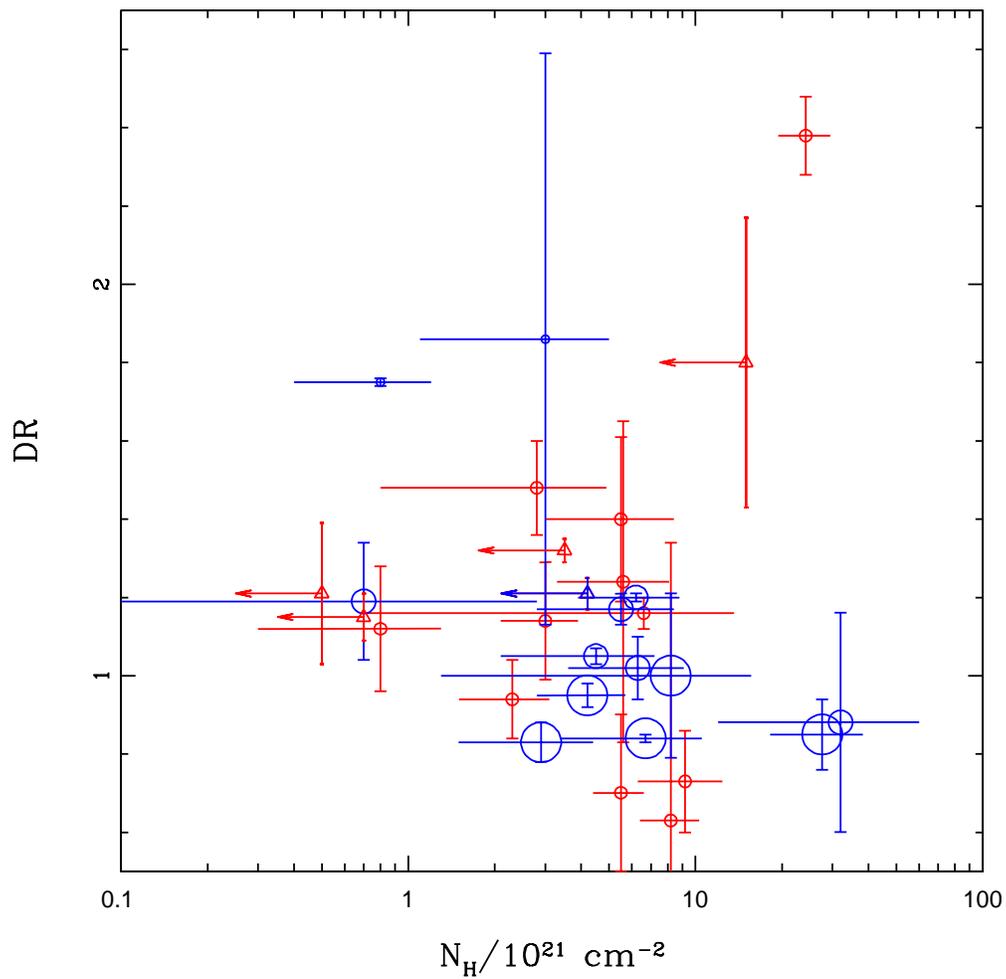}
\caption{Relation between X-ray column density $\mathrm{N_{H,X}}$ and intervening \ion{Mg}{2} DR
for the bursts listed in our sample. The GRBs with the only upper limit of  $\mathrm{N_{H,X}}$ are marked by the open 
triangles and arrows. The red and blue points show the lines-of-sight with single and multiple 
intervening absorption systems, respectively. The size of the blue points is scaled to the number of 
saturated absorbers with $\mathrm{DR}<1.2$. The errorbars correspond to the 1$\sigma$ significance level.\label{fig3}}
\end{figure}

\clearpage

\begin{figure}
\epsscale{.80}
\plotone{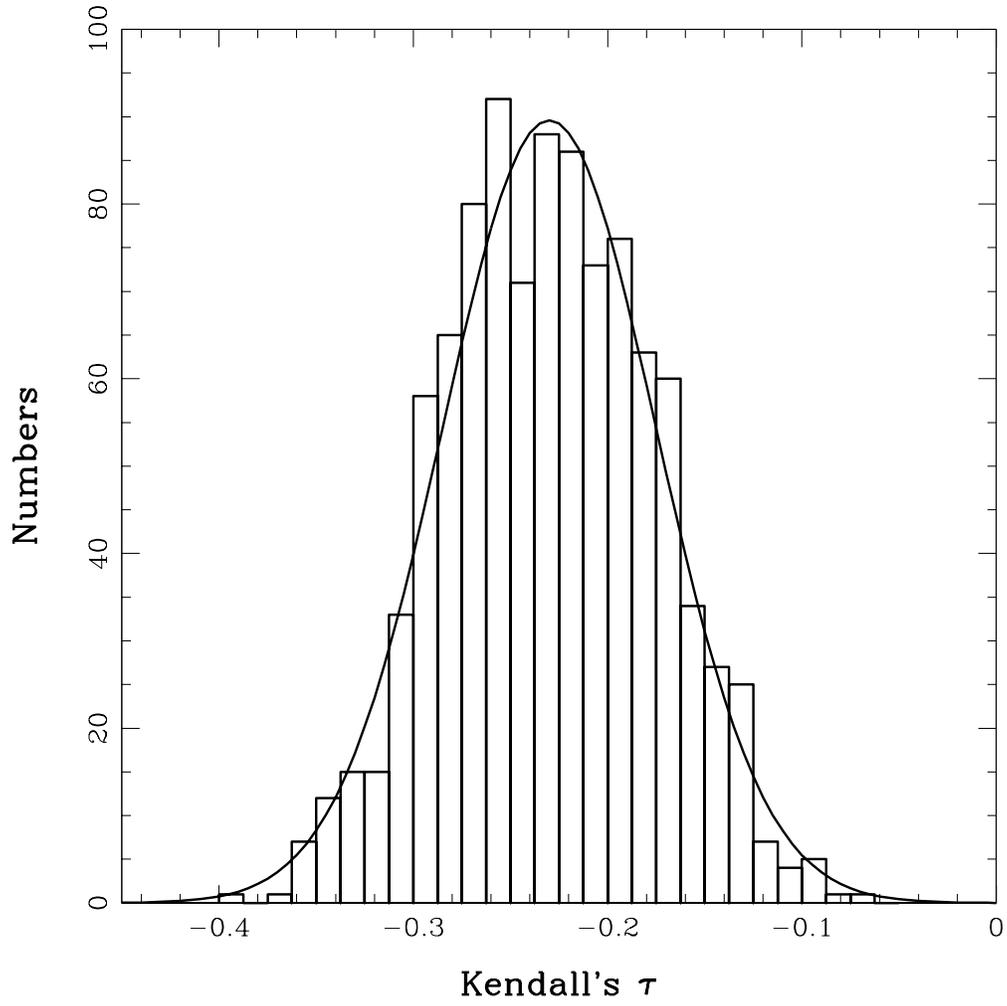}
\caption{Distribution of the simulated Kendall's $\tau$. The best fit Gaussian function is shown by the solid line. \label{fig4}}
\end{figure}

\clearpage

\begin{figure}
\epsscale{.80}
\plotone{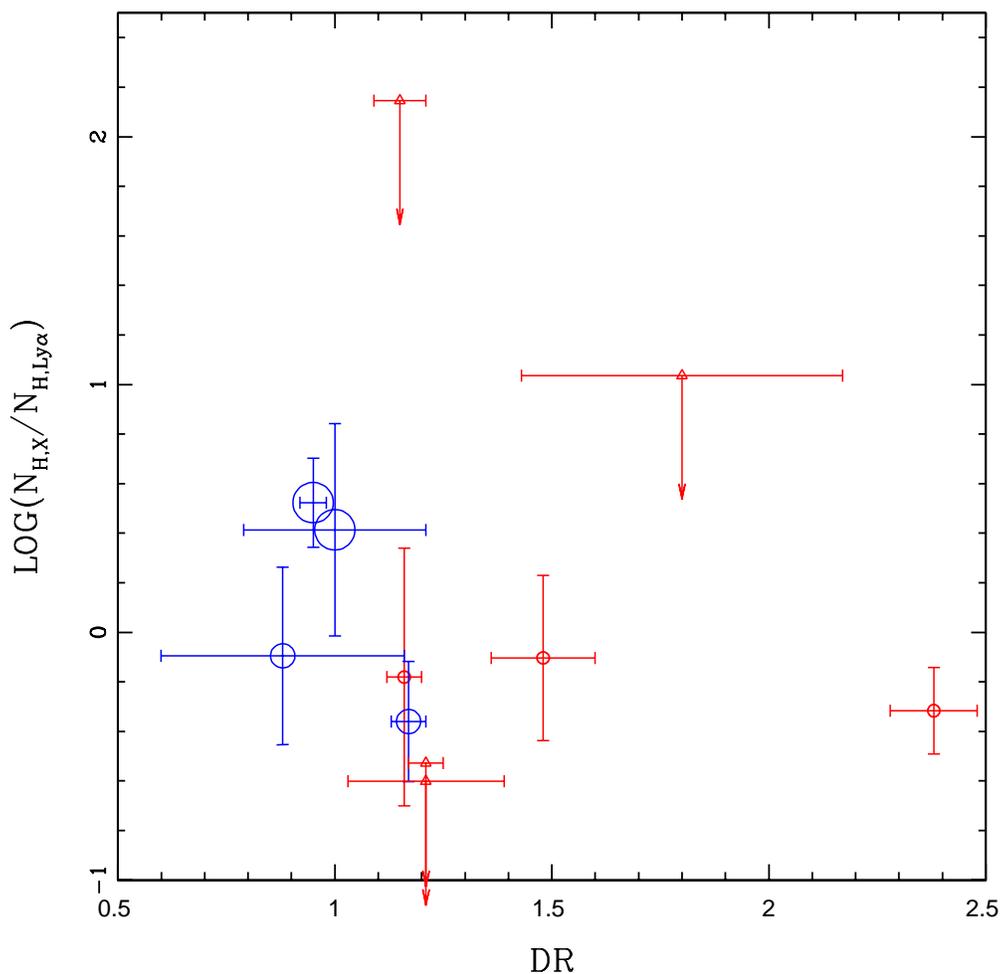}
\caption{\ion{Mg}{2} DR versus the ratio between $\mathrm{N_H}$ measured from the X-ray spectrum and that measured from 
local Lyman-$\alpha$ absorption. Again, the GRBs with only upper limits of $\mathrm{N_{H,X}}$ are marked by the open 
triangles and arrows. Two GRBs, GRB\,060607 and GRB\,050908, are not shown in the diagram because of their extremely
large column density ratios (see text for details). The red and blue points show the lines-of-sight with single and multiple 
intervening absorption systems, respectively. The size of the blue points is scaled to the number of 
saturated absorbers with $\mathrm{DR}<1.2$. The errorbars correspond to the 1$\sigma$ significance level.\label{fig5}}
\end{figure}

%\begin{figure}
%\epsscale{.80}
%\plotone{FeIIvsNH.epsi}
%\caption{\it Top panel:\rm\ the X-ray column density is plotted against the parameter $\mathrm{R_1=EW(MgII\lambda2796)/EW(FeII\lambda2600)}$.
%The dashed horizontal line marks the value of $R_1=1.5$. \it Middle panel:\rm\ the same as top panel but for 
%the parameter $\mathrm{R_2=EW(MgI\lambda2852)/EW(MgII\lambda2796)}$. The dashed line corresponds to $R_2=0.2$.
%\it Bottom panel:\rm\ a comparison of $\mathrm{N_{H,X}}$ distributions between the full sample (histogram by solid line) listed in Table 1 and the 
%sub-sample (shadowed histogram by dashed line) with measured $R_1$. One can see from the comparison that the sub-sample that has large 
%neutral hydrogen column density as indicated by parameters $R_1$ and $R_2$ tends to be biased towards high X-ray column density. \label{fig2}}
%\end{figure}

\clearpage

\clearpage

\begin{table}
\tiny
\begin{center}
\caption{Sample of \it Swift\rm\  GRBs with detection of intervening \ion{Mg}{2} doublet absorption.\label{tbl-1}}
\begin{tabular}{llccccccc}
\tableline\tableline
GRB & $z_{\mathrm{GRB}}$ & $z_{\mathrm{abs}}$ &  EW(2796) & DR  & $\mathrm{N_{H,X}}$ & $\mathrm{N_{HI}}$ & $A_{\mathrm V}$ & Reference\\
    &     &       &   \AA   &     &          $10^{21}\mathrm{cm^{-2}}$  & $10^{21}\mathrm{cm^{-2}}$  & mag  & \\
 (1) & (2) & (3) & (4) & (5) & (6) & (7) & (8) & (9) \\ 
%\multicolumn{1}{c}{$P$\tablenotemark{a}} & $P R_{maj}$ & $P R_{min}$ &
%\multicolumn{1}{c}{$\Theta$\tablenotemark{b}} \\
\tableline
050730  & 3.97    & \bf 1.77317\tablenotemark{*}\rm, 2.25378   &   $0.923\pm0.019$ & $1.17\pm0.04$  &  $5.5^{+2.9}_{-2.7}$ & $12.59\pm2.90$  & \dotfill & 1,5\\
050820  & 2.6147  & 0.6915, \bf 1.4288\tablenotemark{*}\rm, 1.6204, 2.3598\tablenotemark{*}  & $1.203\pm0.023$ & $0.95\pm0.03$  &  $4.2^{+1.5}_{-1.4}$ & $1.26\pm0.29$ & \dotfill & 2,5\\
050908  & 3.339   & 1.548                                & $1.21\pm0.02$   & $1.32\pm0.03$  &  $<3.5$              & $(4,0\pm0.9)\times10^{-4}$ & \dotfill & 2,5\\
050922C & 2.199   & 1.10731                              & $0.532\pm0.031$ & $1.48\pm0.12$  &  $2.8^{+2.1}_{-2.0}$ & $3.55\pm0.82$ & \dotfill & 1,5\\
051111  & 1.549   & 0.82735, \bf 1.18910                 & $2.091\pm0.011$ & $1.20\pm0.01$  &  $6.2^{+2.6}_{-2.3}$ & \dotfill       & \dotfill & 1\\
060418  & 1.49    & \bf 0.60259\tablenotemark{*}\rm, 0.65593, 1.10724, 1.32221  & $1.299\pm0.015$ & $1.05\pm0.02$  &  $4.5^{+2.7}_{-2.4}$ & \dotfill       & \dotfill & 1\\
060502A & 1.515   & 1.044\tablenotemark{*}, 1.078, \bf 1.147\tablenotemark{*}  & $2.39\pm0.12$    & $0.83\pm0.05$  &  $2.9^{+1.5}_{-1.4}$ & \dotfill       & \dotfill & 2\\
060607  & 3.082   & 1.51057, 1.80208, \bf 2.2784\tablenotemark{*}    &  $0.293\pm0.015$       & $1.02\pm0.08$  &  $6.3^{+2.8}_{-2.7}$ & $(8.9\pm0.6)\times10^{-5}$ & \dotfill & 1,5\\
060926  & 3.2     & \bf 1.7954\tablenotemark{*}\rm, 1.8289     &  $3.27\pm0.69$ & $0.88\pm0.28$  &  $32.0^{+28.0}_{-20.0}$ &  $39.8\pm13.8$ & \dotfill & 2,5\\
061007  & 1.261   & 1.065                                & $3.14\pm0.53$  & $0.70\pm0.20$  &  $5.5^{+1.1}_{-1.1}$ & \dotfill       & $0.45^{+0.01}_{-0.01}$ & 2,6\\
070506  & 2.306   & 1.6                                  & $1.92\pm0.04$  & $1.16\pm0.04$  &  $6.6^{+7.0}_{-5.9}$ & $10.0\pm6.9$ & \dotfill & 2,5\\
070611  & 2.039   & 1.297                                & $2.65\pm0.27$  & $1.21\pm0.18$  &  $<0.5$              & $3.2\pm1.5$ & \dotfill & 2,5\\
070802  & 2.45    & \bf 2.0785\tablenotemark{*}\rm, 2.2921\tablenotemark{*}  &  $0.82\pm0.12$ & $1.00\pm0.21$  &  $8.2^{+7.4}_{-6.9}$ & $21.50\pm0.20$ & $1.23^{+0.18}_{-0.16}$ & 2,5,7\\
071003  & 1.604   & \bf0.372\tablenotemark{*}\rm,0.943,1.101   & $2.28\pm0.19$   & $1.19\pm0.15$  &  $0.7^{+2.1}_{-0.6}$ & \dotfill       & \dotfill & 2\\
071031  & 2.6922  & 1.0743, 1.6419, \bf1.9520            & $0.743\pm0.016$   & $1.21\pm0.04$  &  $<4.2$              & $14.1\pm1.6$ & $0.02^{+0.03}_{-0.02}$ & 2,5,7\\
080310  & 2.4272  & 1.6711                               & $0.421\pm0.012$   & $1.15\pm0.06$  &  $<0.7$              & \dotfill & \dotfill & 2\\
080319B & 0.94    & \bf0.5308\rm, 0.5662, 0.7154, 0.7608 & $0.614\pm0.001$     & $1.75\pm0.01$  &  $0.8^{+0.4}_{-0.4}$ & \dotfill       & \dotfill & 2\\
080319C & 1.95    & 0.8104                               & $2.04\pm0.52$ & $1.24\pm0.41$  &  $5.6^{+2.5}_{-2.3}$ & \dotfill       & \dotfill & 2\\
080603A & 1.688   & 1.271\tablenotemark{*}, \bf1.563\tablenotemark{*}   &   $0.77\pm0.01$  & $0.84\pm0.01$  &  $6.7^{+3.8}_{-3.3}$ & \dotfill       & \dotfill & 2\\
080605  & 1.64    & 1.2987                               & $1.08\pm0.11$  & $1.40\pm0.21$  &  $5.5^{+2.9}_{-2.5}$ & \dotfill       & $1.2^{+0.1}_{-0.1}$ & 2\\
080607A & 3.036   & 1.341                                & $3.0\pm0.08$   & $2.38\pm0.10$  &  $24.2^{+5.2}_{-4.8}$ & $50.1\pm17.3$ & $2.33^{+0.46}_{-0.43}$ & 2,5,8\\
081222  & 2.77    & \bf0.8168\rm, 1.0708                 & $0.52\pm0.01$  & $1.86\pm0.73$  &  $3.0^{+2.0}_{-1.9}$ & \dotfill & \dotfill & 2\\
090313  & 3.37    & 1.80\tablenotemark{*}, \bf1.96\tablenotemark{*}    & $0.50\pm0.03$  & $0.85\pm0.09$  &  $27.6^{10.7}_{-9.4}$ & \dotfill & $0.42^{+0.06}_{-0.05}$ & 3,7\\
091208B & 1.063   & 0.784                                & $0.65\pm0.43$  &  $0.63\pm0.71$  &  $8.2^{+2.1}_{-1.8}$ & \dotfill & \dotfill & 2\\
100219A & 4.6667  & 2.18                                 & $0.90\pm0.08$  & $1.80\pm0.37$  &  $<15.0$             & $1.38\pm0.16$ & \dotfill & 4,5\\
100814A & 1.44    & 1.1574                               & $0.426\pm0.04$ &  $1.12\pm0.16$  &  $0.8^{+0.5}_{-0.5}$ & \dotfill & \dotfill & 2\\
100901A & 1.408   & 1.314                                & $1.74\pm0.17$  &  $1.14\pm0.15$  &  $3.0^{+0.9}_{-0.9}$ & \dotfill &\dotfill &  2\\
100906A & 1.64    & 0.994                                & $0.87\pm0.10$ &  $0.73\pm0.13$  &  $9.2^{+3.2}_{-2.9}$ & \dotfill & \dotfill & 2\\
110918A & 0.982   & 0.877                                & $2.65\pm0.20$   & $0.94\pm0.10$  &  $2.3^{+0.8}_{-0.8}$ & \dotfill & \dotfill & 2\\
\tableline
\end{tabular}
%% Any table notes must follow the \end{tabular} command.
%\tablenotetext{a}{The \ion{Mg}{2} doublet is not distinguishable because of the strong line blend. The $R_1$ and $R_2$ parameters are 
%calculated by assuming the corresponding \ion{Mg}{2} DR=1.}
%\tablenotetext{c}{Another sample footnote for table~\ref{tbl-2}}
%\tablecomments{We can also attach a long-ish paragraph of explanatory
%material to a table.}
\tablecomments{References in the last column: 1-Tejos et al. (2009); 2-Cucchiara et al. (2012); 3-de Ugarte Posigo et al. (2010); 
4-Thone et al. (2013);  5-Fynbo et al. (2009); 6-Schady et al. (2010); 7-Greiner et al. (2011); 8-Zafar et al. (2011)}

\end{center}
\end{table}

%% If the table is more than one page long, the width of the table can vary
%% from page to page when the default \tablewidth is used, as below.  The
%% individual table widths for each page will be written to the log file; a
%% maximum tablewidth for the table can be computed from these values.
%% The \tablewidth argument can then be reset and the file reprocessed, so
%% that the table is of uniform width throughout. Try getting the widths
%% from the log file and changing the \tablewidth parameter to see how
%% adjusting this value affects table formatting.

%% The \dataset{} macro has also been applied to a few of the objects to
%% show how many observations can be tagged in a table.

%% Tables may also be prepared as separate files. See the accompanying
%% sample file table.tex for an example of an external table file.
%% To include an external file in your main document, use the \input
%% command. Uncomment the line below to include table.tex in this
%% sample file. (Note that you will need to comment out the \documentclass,
%% \begin{document}, and \end{document} commands from table.tex if you want
%% to include it in this document.)


%% The following command ends your manuscript. LaTeX will ignore any text


\begin{thebibliography}{}
\bibitem[Behar et al. 2011]{beh01} Behar, E., Dado, S., Dar, A., \& Laor, A. 2011, \apj, 734, 26
\bibitem[Butler et al. 2003]{but03} Butler, N. R., Marshall, H. L., Ricker, G. R., Vanderspek, R. K., Ford, P. G., Crew, G. B., Lamb, D. Q., \& Jernigan, J. G. 2003, \apj, 597, 1010
\bibitem[Campana et al. 2006]{cam06}  Campana, S., et al. 2006, \aap, 449, 61
\bibitem[Campana et al. 2007]{cam07} Campana, S., et al. 2007, \apjl, 654, 17
\bibitem[Campana et al. 2010]{cam10}  Campana, S., Thone, C. C., de Ugarte Postigo, A., Tagliaferri, G., Moretti, A., \& Covino, S. 2010, \mnras, 402, 2429
\bibitem[Campana et al. 2012]{cam12} Campana, S., Salvaterra, R., Melandri, A., Vergani, S. D., Covino, S., D'Avanzo, P., Fugazza, D., Ghisellini, G., et al. 2012, \mnras, 421, 1697
\bibitem[Covino et al. 2013]{cov13} Covino, S., et al. 2013, astro-ph/arXiv:1303.4743 
\bibitem[Cucchiara et al. 2012]{cuc12} Cucchiara, A., et al. 2012, astro-ph/arXiv:1211.6528 
\bibitem[Churchill et al. 2000]{chu00} 	Churchill, C. W., Mellon, R. R., Charlton, J. C., Jannuzi, B. T., Kirhakos, S., Steidel, C. C., \& Schneider, D. P. 2000, \apj, 543, 577
%\bibitem[D'Eila et al. 2010]{dei10} D'Elia, V., et al. 2010, \aap, 523, 36
\bibitem[de Ugarte Posigo et al. 2010]{deu10} de Ugarte Posigo, A., et al. 2010, \aap, 513, 42 
\bibitem[Evans et al. 2009]{eva09} Evans, P. A., et al. 2009, \mnras, 397, 1177
\bibitem[Fynbo et al. 2009]{fyn09} Fynbo, J. P. U., et al. 2009, \apjs, 185, 526
\bibitem[Fruchter et al. 2001]{fru01} Fruchter, A., Krolik, J. H., \& Rhoads, J. E. 2001, \apj, 563, 597
\bibitem[Galama \& Wijers 2001]{gaw01} Galama, T. J., \& Wijers, R. A. M. J. 2001, \apjl, 549, 209
\bibitem[Gendre et al. 2007]{gen07} Gendre, B., Galli, A., Corsi, A., Klotz, A., Piro, L., Stratta, G., Boer, M., \& Damerdji, Y. 2007, \aap, 462, 565
\bibitem[Greiner et al. 2011]{gre11} Greiner, J., et al. 2011, \aap, 526, 30
\bibitem[Gehrels et al. 2004]{geh04} Gehrels, N., et al. 2004, \apj, 611, 1005
\bibitem[Gupta et al. 2009]{gup09} Gupta, N., Srianand, R., Petitjean, P., Noterdaeme, P., \& Saikia, D. J. 2009, \mnras, 398, 201 
\bibitem[Hjorth \& Bloom 2011]{hjb11}Hjorth, J., \&  Bloom, J. S. 2011, astrp-ph/arXiv:1104.2274
\bibitem[Kruhler et al. 2011]{kru11} Kruhler, T., et al. 2011, \aap, 543, 108
\bibitem[Menard \& Chelouche 2009]{mec09} Menard, B., \& Chelouche, D. 2009, \mnras, 393, 808
\bibitem[Meszaros \& Rees 1997]{mer07} Meszaros, P., \& Rees, M. J. 1997, \apjl, 482, 29
\bibitem[Perna \& Lazzati 2001]{pel01} Perna, R., \& Lazzati, D. 2001, \apj, 580, 261
\bibitem[Perna et al. 2003]{per03} Perna, R., Lazzati, D., \& Fiore, F. 2003, \apj, 585, 775
\bibitem[Petitjean \& Bergeron 1990]{peb90} Petitjean, P., \& Bergeron, J. 1990, \aap, 231, 309
\bibitem[Prochaska et al. 2007]{pro07} Prochaska, J. X., et al. 2007, \apjs, 168, 231
\bibitem[Prochter et al. 2006]{pro06} Prochter, G. E., et al. 2006, \apj, 648, 93
\bibitem[Rao \& Turnshek 2000]{rat00} Rao, S. M., \& Turnshek, D. A. 2000, \apjs, 130, 1
\bibitem[Rigby et al. 2002]{rig02} Rigby, J. R., Charlton, J. C., \& Churchill, C. W. 2002, \apj, 565, 743
\bibitem[Salvaterra et al. 2009]{sal09} Salvaterra, R., Della Valle, M., Campana, S., Chincarini, G., Covino, S., D'Avanzo, P., Fernandez-Soto, A., Guidorzi, C., et al.
2009, \nat, 461, 1258
\bibitem[Sari et al. 1998]{sar98} Sari, R. Piran, T., \& Narayan, R. 1998, \apjl, 497, 17 
\bibitem[Savaglio \& Fall 2004]{saf04} Savaglio, S., \& Fall, S. M. 2004, \apj, 614, 293
\bibitem[Savaglio et al. 2003]{sav03} Savaglio, S. Fall, S. M., \& Fiore, F. 2003, \apj, 585, 638
\bibitem[Schady et al. 2007]{sch07} Schady, P., et al. 2007, \mnras, 377, 273
\bibitem[Schady et al. 2010]{sch10} Schady, P., et al. 2010, \mnras, 401, 2773
\bibitem[Starling et al. 2013]{sta13} Starling, R. L. C., Willingale, R., Tanvir, N. R., Scott, A. E., Wiersema, K., O'Brien, P. T., Levan, A. J., \& Stewart, G. C. 2013,
astro-ph/arXiv:1303.0844
\bibitem[Starling et al. 2007]{sta07} Starling, R. L. C., Wijers, R. A. M. J., Wiersema, K., Rol, E., Curran, P. A., Kouveliotou, C., van der Horst, A. J., \& Heemskerk, M. H. M.
2007, \apj, 661, 787
\bibitem[Stratta et al. 2004]{str04} Stratta, G., Fiore, F., Antonelli, L. A., Piro, L., \& De Pasquale, M. 2004, \apj, 608, 846
\bibitem[Tanvir et al. 2009]{tan09}Tanvir, N. R., Fox, D. B., Levan, A. J., Berger, E., Wiersema, K., Fynbo, J. P. U., Cucchiara, A., \& Kruhler, T., et al. 2009, \nat,
461, 1254
\bibitem[Tejos et al. 2009]{tej09} Tejos, N., Lopez, S., Prochaska, J. X., Bloom, J. S., Chen, H. W., Dessauges-Zavadsky, M., \& Maureira, M. J. 2009, \apj, 706, 1309
\bibitem[Thone et al. 2013]{tho13} Thone, C. C., et al. 2013, \mnras, 428, 3590
\bibitem[Vergani et al. 2009]{ver09} Vergani, S. D., Petitjean, P., Ledoux, C., Vreeswijk, P., Smette, A., \& Meurs, E. J. A. 2009, \aap, 503, 771
\bibitem[Watson et al. 2007]{wat07}Watson, D., Hjorth, J., Fynbo, J. P. U., Jakobsson, P., Foley, S., Sollerman, J., \& Wijers, R. A. M. J. 2007, \apjl, 660, 101
\bibitem[Watson \& Jakobsson 2012]{waj12} Watson, D., \& Jakobsson, P. 2012, \apj, 754, 89
\bibitem[Watson et al. 2013]{wat13} Watson, D., et al. 2013, \apj, 768, 23
\bibitem[Welty et al. 2012]{wel12} Welty, D. E., Xue, R., \& Wong, T. 2012, \apj, 745, 173
\bibitem[Woosley \& Bloom 2006]{wob06} Woosley, S. E., \& Bloom, J. S. 2006, \araa, 44, 507
\bibitem[Zafar et al. 2011]{zaf11} Zafar, T., Watson, D., Fynbo, J. P. U., Malesani, D., Jakobsson, P., \& de Ugarte Postigo, A. 2011, \aap, 532, 143
\end{thebibliography}
\end{document}